\documentclass{kapproc}
\begin{document}

\articletitle{Understanding the IMF}

\author{Richard B. Larson}

\affil{Department of Astronomy, Yale University\\
Box 208101, New Haven, CT 06520-8101, USA}

\email{larson@astro.yale.edu}

\begin{abstract}
It is suggested that the thermal physics of star-forming clouds may play a more important role than has usually been recognized in the origin of the stellar IMF and in determining a characteristic mass scale.  The importance of the thermal physics has been clearly demonstrated for the formation of the first stars in the universe, where it is well understood and results in cooling to a characteristic minimum temperature at a preferred density, and hence in a characteristic scale for fragmentation.  In present-day star-forming clouds, an analogous situation may exist in that at low densities the temperature is expected to decrease with increasing density, reaching a minimum when the gas becomes thermally coupled to the dust and then rising slowly at higher densities.  A minimum temperature of about 5~K is predicted to occur at a density of the order of $10^{-18}$\,g\,cm$^{-3}$, and at this point the Jeans mass is about 0.3 solar masses, similar to the mass at which the IMF peaks.  If most of the fragmentation in star-forming clouds occurs in filaments, as is suggested by many simulations as well as by observations, fragmentation seems likely to occur preferentially at the density where the temperature reaches a minimum, and the Jeans mass at this point may then determine a characteristic scale for fragmentation and hence a preferred stellar mass.
\end{abstract}

\section*{Introduction}

   As we have seen at this meeting, much evidence now supports a general form for the stellar IMF which is similar to the original Salpeter power law at masses above one solar mass, but which flattens at lower masses and peaks at a few tenths of a solar mass when expressed in terms of the number of stars per unit logarithmic mass interval (Scalo 1986, 1998; Kroupa 2002; Larson 2003a).  Nearly all studies of the lower end of the IMF have shown that the IMF declines in the brown-dwarf regime, making it clear that the IMF is a peaked function.  Nature thus makes stars with a preferred mass that is between one-tenth and one solar mass.  The amount of mass that goes into stars in each logarithmic mass interval is also a broadly peaked function that has a maximum at about 0.5 solar masses, according to the approximation suggested by Kroupa (2002).  Thus, in terms of where most of the mass goes, there is a characteristic stellar mass that is of the order of half of a solar mass.  This is probably the most fundamental fact about star formation that needs to be explained by any theory of how stars form: some feature of the physics of star formation must result in a characteristic stellar mass a little below one solar mass.

   A further remarkable fact for which we have seen much evidence here is that the IMF shows a considerable degree of universality: very similar, and often indistinguishable, results for the IMF are found in many different star-forming environments in our Galaxy and other nearby galaxies, and no clear dependence has been found on any plausibly relevant astrophysical parameters such as metallicity.  Thus, not only must some aspect of the fundamental physics of star formation lead to a preferred mass scale, but it must operate in a relatively universal way that is only weakly dependent on the environment and on most astrophysical parameters.  This fact poses a clear challenge to the theory of star formation.

\section{The Role of Cloud Fragmentation and the Jeans Mass}

   Recent millimeter-wavelength surveys of star-forming clouds have revealed the existence of many small dense clumps that have masses extending well down into the brown-dwarf regime, and that appear likely to be direct stellar progenitors (e.g., Motte et al 1998; Motte \& Andr\'e 2001a,b).  Part of the evidence that these small dense clumps or `cloud cores' may be direct stellar progenitors is the fact that their mass spectrum is similar to the stellar IMF at masses below a few solar masses (Luhman \& Rieke 1999; Motte \& Andr\'e 2001a,b), suggesting that stars in this mass range may gain their masses directly from those of the cloud cores in which they form.  Other authors have found similar results, but with core masses that are systematically somewhat larger (Johnstone et al 2000, 2001); these results could still be consistent with the direct collapse of these cores into stars but with a somewhat lower efficiency of star formation.  If low-mass stars gain their masses from those of the prestellar cloud cores in which they form, the problem of understanding stellar masses then becomes, to a large extent, the problem of understanding cloud fragmentation processes, i.e. of understanding how the material in a star-forming cloud becomes divided up into individual star-forming units.

   Perhaps even more direct evidence that low-mass stars owe their characteristic mass to cloud fragmentation processes is provided by the fact that stars typically form in small clusters of a few hundred stars in which the efficiency of star formation is moderately high, of the order of 25 to 30 percent (Lada \& Lada 2003).  This means that the average stellar mass is determined to within a factor of 3 or 4 just by the number of stars that form in each cluster-forming region, since the average mass is equal to the total stellar mass divided by the number of stars, and this is equal within a factor of 3 or 4 to the total mass of the cluster-forming region divided by the number of stars formed in it.  The problem of understanding the preferred stellar mass is then basically one of understanding the number of star-forming cloud cores that form in such a region.  That is, the problem is one of understanding the typical scale of fragmentation -- does most of the mass go into a few large objects or many small ones?

   The most classical type of fragmentation scale is the well-known Jeans scale based on balancing gravity against thermal pressure, which has been known for more than 100 years (Jeans 1902, 1929).  Although the original derivation by Jeans of a minimum length scale for fragmentation in an infinite uniform medium was not self-consistent, rigorous stability analyses have yielded dimensionally equivalent results for a variety of equilibrium configurations, including sheets, filaments, and spheres (Larson 1985, 2003b).  Many simulations of cloud collapse and fragmentation that include gravity and thermal pressure have also shown a clear imprint of the Jeans scale on the results: the number of star-forming cloud cores formed is always comparable to the number of Jeans masses present initially (e.g., Larson 1978; Monaghan \& Lattanzio 1991; Klessen 2001; Bate, Bonnell, \& Bromm 2003).  As long as gravity is strong enough to cause collapse to occur, the scale of fragmentation is not expected to be greatly altered by the presence of rotation or magnetic fields (Larson 1985, 2003b).  Simulations of turbulent fragmenting clouds have also shown that the amount of fragmentation that occurs is not strongly dependent on the way in which turbulence is introduced, or even on whether turbulence is initially present at all; the number of fragments formed always remains similar to the number of Jeans masses present initially, although fragment masses may be somewhat reduced by compression occurring during the collapse (Klessen 2001; Bonnell \& Bate 2002; Bate et al 2003).  The effect of a magnetic field is less clear, but simulations of MHD turbulence typically show that that the nature of the turbulence is not qualitatively altered by the presence of a magnetic field; similar filamentary and clumpy structures are still seen (Mac Low \& Klessen 2004).

   Thus, while the relevance of the Jeans scale has not been established in all circumstances, many theoretical and numerical results have suggested that it has wide applicability.  Its relevance to star formation might not at first seem obvious, given that during the early stages of the process gravity is opposed mainly by magnetic fields and turbulence, while the Jeans criterion involves only gravity and thermal pressure (Mac Low \& Klessen 2004).  However, the early evolution of star-forming clouds is expected to be characterized, to a considerable extent, by the loss of magnetic flux and the dissipation of turbulence in the densest contracting core regions, leaving thermal pressure eventually as the main force opposing gravity in the small dense prestellar cores that form.  Thus star formation in effect gets rid of most of the initial magnetic field and turbulence, as well as angular momentum, at a relatively early stage.  As would be expected on this basis, observed prestellar cores show a rough balance between gravity and thermal pressure, with a minor contribution from turbulence and a contribution from magnetic fields that may be comparable to thermal pressure but probably is not dominant.  The thermal physics must then play an important role in the later stages of the star formation process and in the processes that determine stellar masses and the IMF.

\section{Importance of the Thermal Physics of Star-Forming Clouds}

   If the Jeans criterion indeed plays a significant role in determining stellar masses, it is important to understand the thermal behavior of star-forming clouds because the Jeans mass depends strongly on the temperature, varying either as $T^{3/2}\rho^{-1/2}$ or as $T^2P^{-1/2}$, depending on whether the density or the pressure of the medium is specified.  Most of the simulations of cloud collapse and fragmentation that have been made in the past few decades have adopted a simple isothermal equation of state, and the results show, as was noted above, that fragmentation during isothermal collapse is limited to producing a number of fragments that is typically comparable to the number of Jeans masses present in the initial cloud.  This means that the scale of fragmentation during isothermal collapse is largely determined by the initial conditions, or by the conditions existing when isothermal collapse begins.  In the simulations the initial conditions can be chosen arbitrarily, but how does nature choose the initial conditions?  Why should nature prefer initial conditions for isothermal collapse that yield a Jeans mass of the order of one solar mass?

   One way of explaining a universal mass scale might be in terms of some universality in the properties of star-forming molecular clouds or their internal turbulence, for example a characteristic turbulent ram pressure that translates into a preferred Jeans mass (Larson 1996).  The turbulent pressures in nearby star-forming clouds have about the right order of magnitude for such an explanation to seem feasible (Larson 2003a).  However, beyond the local region, star-forming clouds and environments in our Galaxy and others vary so widely in their properties, and turbulence is such a variable and poorly defined phenomenon, that this view does not clearly offer an appealing explanation for a quasi-universal IMF.

   Another possibility, not widely explored until now, is that there is some universality in the internal physics of star-forming clouds, for example in their thermal physics, that results in a preferred mass scale for fragmentation.  This might be the case if the gas does not remain isothermal during the collapse, since variations in the temperature could then have an important effect on fragmentation, given the strong sensitivity of the Jeans mass to temperature.  For example, if significant cooling occurs during the early stages of collapse, this can greatly increase the amount of fragmentation that occurs during these stages (Monaghan \& Lattanzio 1991).  The temperatures of star-forming clouds are controlled by processes of atomic and molecular physics that should operate in much the same way everywhere, and this might result in a mass scale for fragmentation that depends only on atomic-scale  physics and therefore is relatively universal.

   One problem of star formation in which the thermal physics is well understood and has a clear impact on the scale of fragmentation is the problem of the formation of the first stars in the universe, before any heavy elements had been produced.  This topic is reviewed briefly in the next section as an example of the possible importance of the thermal physics for stellar masses and the IMF.

\section{An Example: The Masses of the First Stars}

   In contrast to present-day star formation where the thermal physics is complicated and involves many types of processes and particles, the thermal physics of the first star-forming clouds is relatively simple; because of the absence of any heavy elements, it involves only hydrogen molecules that control the temperature by their infrared line emission.  As reviewed by Bromm \& Larson (2004), calculations by several groups made with varying assumptions have converged on a consistent picture of the thermal behavior of the first star forming clouds that is largely independent of many of the details, including the cosmological initial conditions, and is determined basically by the physics of the H$_2$ molecule.  A parameter exploration by Bromm, Coppi, \& Larson (2002) also showed a robust thermal behavior that was largely independent of the parameters varied.

   The first star-forming systems or primitive `protogalaxies' are predicted to form at redshifts of the order of 20 to 30, and because of the absence at first of any significant coolants, the gas collapses adiabatically in each small dark halo and forms a rotating, flattened, possibly disk-like configuration.  Compressional heating during this collapse raises the gas temperature to above 1000~K, and this increases the rate of formation of H$_2$ molecules by ion chemistry, producing a molecular abundance fraction of about $10^{-3}$.  The hydrogen molecules then become an important coolant via their infrared line emission, and this causes the temperature of the denser gas to fall again to a minimum value of about 200~K that is determined by the level spacing of the H$_2$ molecules.  At about the same time, as the gas density approaches $10^4$\,cm$^{-3}$, the upper levels of the H$_2$ molecules become thermalized, and this reduces the density dependence of the cooling rate so that the cooling time no longer decreases with increasing density.  The cooling time then becomes longer than the free-fall time, and the densest clumps approach a rough balance between pressure and gravity.  They then nearly satisfy the Bonnor-Ebert criterion for a marginally stable isothermal sphere, which is dimensionally equivalent to the Jeans criterion.  The masses of these slowly contracting clumps therefore become essentially equal to the Jeans or `Bonnor-Ebert' mass at a characteristic temperature and density of about 200~K and $10^4$\,cm$^{-3}$.  Calculations by several groups (Abel, Bryan, \& Norman 2000, 2002; Bromm, Coppi, \& Larson 1999, 2002; Nakamura \& Umemura 2001, 2002) have all shown convergence into a similar regime of temperature and density.  In this preferred regime, the Bonnor-Ebert mass is several hundred solar masses, and as might be expected, all of the simulations have yielded clump masses of this order, that is, between a few hundred and a thousand solar masses.

   The final evolution of these clumps has not yet been determined, but none of the simulations has yet shown any tendency for them to fragment further when the calculations are followed with high resolution to much higher densities.  This is perhaps not too surprising, given that the temperature rises slowly with increasing density and the contraction of the clumps is slowed by inefficient cooling, making it seem unlikely that further fragmentation will occur, except possibly for the formation of binaries.  Therefore it seems likely that a star with a mass of at least a hundred solar masses will form, unless feedback effects strongly inhibit accretion during the later stages.  Omukai \& Palla (2003) have studied the accretion problem with spherical symmetry, including all of the relevant radiative effects, and they have found that if the accretion rate is not too high, accretion can continue up to a mass of at least 300 solar masses.  High-resolution calculations of the accretion phase by Bromm \& Loeb (2004) have yielded accretion rates that according to Omukai \& Palla (2003) are in the range that can allow accretion to continue up to a maximum mass that could be as large as 500 solar masses (Bromm 2005).  While this is not yet a final answer because radiative effects have not yet been included in a realistic three-dimensional calculation, it seems very likely that the first stars were indeed massive objects, quite possibly more massive than 100 solar masses.  If this conclusion is correct, this large mass scale can be attributed basically to the physics of the H$_2$ molecule, and it depends only secondarily on other factors such as the details of the cosmological initial conditions.

\section{Thermal Physics of Present-Day Star-Forming Clouds}

   As was noted above, the thermal behavior of the first collapsing clouds is expected to be characterized by an early phase of rapid cooling by H$_2$ line emission to a minimum temperature of $\sim 200$~K, followed by a phase of slower contraction during which temperature slowly rises again.  A qualitatively similar situation is expected to occur in present-day star-forming clouds: as can be seen in Figure 2 of Larson (1985), observations and theory both show that at low densities the temperature decreases with increasing density, in this case because of cooling by line emission from C$^+$ ions and O atoms, while at high densities, the temperature is predicted to rise slowly with increasing density; this occurs when the gas becomes thermally coupled to the dust grains, which then control the temperature by their far-infrared thermal emission.  Between these two thermal regimes, the temperature reaches a predicted minimum value of about 5~K at a density of the order of $10^{-18}$\,g\,cm$^{-3}$, at which point the Jeans mass is a few tenths of a solar mass.  Does this expected thermal behavior play a role in determining the resulting stellar IMF?

   The variation of temperature with density described above can be approximated at low densities by a polytropic equation of state \hbox{$P \propto \rho^{\gamma}$} with an exponent of $\gamma \sim 0.7$, and at high densities by a polytropic equation of state with an exponent of $\gamma \sim 1.1$.  The collapse and fragmentation of turbulent clouds with various assumed values of $\gamma$ has been simulated by Li, Klessen, \& Mac Low (2003), and they found that such a variation in $\gamma$ can make a large difference to the amount of fragmentation that occurs: in particular, they found that, for the same initial conditions and the same treatment of turbulence, a simulation with $\gamma = 0.7$ produced about 380 bound condensations, while one with $\gamma = 1.1$ produced only about 20 bound objects, a factor of 20 fewer.  This dramatic reduction in the amount of fragmentation that occurs when $\gamma$ is increased from 0.7 to 1.1 suggests that in real star-forming clouds, much more fragmentation will occur during the low-density phase of collapse when the effective value of $\gamma$ is close to 0.7 than during the high-density phase when $\gamma$ is closer to 1.1.  Fragmentation may then almost come to a halt at the transition density where the temperature reaches a minimum and the Jeans mass is a few tenths of a solar mass.  If this is indeed the case, and if fragmentation and the resulting mass spectrum are really so sensitive to the thermal physics, it is important to understand as accurately as possible the detailed thermal behavior of collapsing and fragmenting clouds.

   The observational and theoretical results reviewed by Larson (1985) were taken from early work that included predictions by Larson (1973) of the temperature-density relation at the higher densities, but more recent work has mostly yielded similar results.  In the low-density regime, the work of Koyama \& Inutsuka (2000), which assumes that heating is due to the photoelectric effect rather than cosmic rays, as had been assumed in earlier work, yields a similar predicted decrease of temperature with increasing density.  The observations compiled by Myers (1978) and plotted in Figure 2 of Larson (1985) had suggested temperatures rising again toward the high end of this regime, but these observations referred mostly to relatively warm and massive cloud cores and not to the small, dense, cold cores in which low-mass stars form; as reviewed by Evans (1999), these low-mass cores have much lower temperatures that are typically only about 8.5~K at a typical density of $10^{-19}$\,g\,cm$^{-3}$, and this value is consistent with a continuation of the decreasing trend seen at lower densities, and with the continuing validity of a polytropic equation of state with $\gamma \sim 0.7$ up to a density of at least $10^{-19}$\,g\,cm$^{-3}$.

   At much higher densities, where the gas becomes thermally coupled to the dust grains, few temperature measurements exist because most of the molecules freeze out onto the grains, but most of the available theoretical predictions (Larson 1973; Low \& Lynden-Bell 1976; Masunaga \& Inutsuka 2000) agree well concerning the expected temperature-density relation, and they are consistent with an approximate power-law dependence with an equivalent $\gamma$ of about 1.1.  All of the theoretical predictions mentioned in this section can be fitted to within about $\pm 0.1$~dex by the following approximation consisting of two power laws:

$$T=4.4\,\rho_{-18}^{-0.275}\,{\rm K},~~\rho < 10^{-18}\,{\rm g\,cm}^{-3}$$
\vskip -8pt
$$T=4.4\,\rho_{-18}^{+0.075}\,{\rm K},~~\rho > 10^{-18}\,{\rm g\,cm}^{-3}$$
\vskip 6pt

\noindent where $\rho_{-18}$ is the density in units of $10^{-18}$\,g\,cm$^{-3}$.  This approximation to the equation of state, in which the value of $\gamma$ changes from 0.725 to 1.075 at a density of $10^{-18}$\,g\,cm$^{-3}$, is valid for densities between about $10^{-22}$\,g\,cm$^{-3}$ and $10^{-13}$\,g\,cm$^{-3}$.  The actual predicted temperature minimum is somewhat smoothed out compared with this simple two-part approximation, and the predicted minimum temperature is actually about 5~K at a density of about $2 \times 10^{-18}$\,g\,cm$^{-3}$.  The minimum temperature attained in real clouds is somewhat uncertain because observations have not yet confirmed the predicted very low values; such cold gas would be very difficult to observe, but various efforts to model the observations have suggested central temperatures between 6~K and 10~K for the densest observed prestellar cores, whose peak densities may approach $10^{-17}$\,g\,cm$^{-3}$. (e.g., Zucconi et al 2001; Evans et al 2001; Tafalla et al 2004).  The temperature minimum may therefore in reality be shallower and more smoothed out than the predicted one, but the above approximation should still be valid for densities below $10^{-19}$\,g\,cm$^{-3}$ or well above $10^{-17}$\,g\,cm$^{-3}$.

\section{The Fragmentation of Filaments}

   To explore the effect of an equation of state in which $\gamma$ changes from about 0.7 at low densities to about 1.1 at high densities, Jappsen et al (2003, 2005) have made calculations of the collapse and fragmentation of turbulent clouds, which are similar to the those of Li et al (2003) except that $\gamma$ is assumed to change from 0.7 to 1.1 at some critical density $\rho_{\rm crit}$ (see also Klessen 2005, this conference).  The value of $\rho_{\rm crit}$ is then varied to test the effect of this parameter on the mass spectrum of the star-forming condensations that form.  The results show a clear dependence of the mass spectrum on $\rho_{\rm crit}$ in the expected sense that as $\rho_{\rm crit}$ is increased and the Jeans mass at the point of minimum temperature is thereby decreased, the number of bound condensations increases and their median mass decreases.  This result shows that the detailed temperature-density relation in a collapsing cloud can play an important role in determining the mass spectrum of the fragments that form.  These results also show that the number of fragments increases with increasing numerical resolution, which indicates that very high resolution is needed to obtain reliable results for the higher values of $\rho_{\rm crit}$.

   A striking feature of these results is the prominence of filamentary structure in the simulated collapsing clouds, and the fact that nearly all of the bound condensations form in filaments.  Filamentary structure is, in fact, a very common feature of simulations of collapse and fragmentation (Monaghan \& Lattanzio 1991; Klessen \& Burkert 2001; Bonnell \& Bate 2002; Bate et al 2003).  Many observed star-forming clouds also exhibit filamentary structure, and together with the evidence that much of the star formation in these clouds occurs in filaments (Schneider \& Elmegreen 1979; Larson 1985; Curry 2002; Hartmann 2002), this suggests that the formation and fragmentation of filaments may be an important mode of star formation quite generally.  If most of the fragmentation that leads to star formation occurs in filaments, this may offer an explanation for the fact that the amount of fragmentation that was found by Li et al (2003) depends so strongly on the assumed value of $\gamma$, especially for values of $\gamma$ near unity.  This sensitivity may result from the fact that $\gamma = 1$ is a critical value for the collapse of cylinders: for $\gamma < 1$, a cylinder can collapse indefinitely toward its axis and fragment indefinitely into many very small objects, while for $\gamma > 1$, this is not possible because pressure then increases faster than gravity and stops the collapse at a finite maximum density, where the Jeans mass has a finite minimum value.  More fragmentation would then be expected to occur when $\gamma < 1$, as was indeed found by Li et al (2003).
 
   The evolution of collapsing configurations can often be approximated by similarity solutions (Larson 2003b), and further insight into the collapse and fragmentation of filaments can be obtained from the similarity solutions derived by Kawachi \& Hanawa (1998) for the collapse of cylinders with a polytropic equation of state.  These authors found that the existence of such solutions depends on the value of $\gamma$: similarity solutions exist for $\gamma < 1$ but not for $\gamma > 1$.  For the solutions with $\gamma < 1$, the collapse becomes slower and slower as $\gamma$ approaches unity from below, asymptotically coming to a halt when $\gamma = 1$.  This result shows in a particularly clear way that $\gamma = 1$ is a critical value for the collapse of filaments.  Kawachi \& Hanawa (1998) suggested that the slow collapse predicted to occur for values of $\gamma$ approaching unity will in reality cause a filament to fragment into clumps, because the timescale for fragmentation then becomes shorter than the timescale for collapse toward the axis.  If in real clouds the effective value of $\gamma$ increases with increasing density as the collapse proceeds, as is expected from the thermal behavior discussed above, fragmentation might then be favored to occur at the density where $\gamma$ reaches unity, i.e. at the density where the temperature reaches a minimum.  As was noted earlier, the Jeans mass at the density where the temperature attains its minimum value is predicted to be about 0.3 solar masses, coincidentally close to the mass at which the stellar IMF peaks.  This similarity suggests that filament fragmentation with an increasing polytropic exponent may play an important role in the origin of the stellar IMF and its characteristic mass scale.  Further calculations to test this hypothesis are currently under way, and some first results have been reported at this conference by Klessen (2005).

\section{Summary}

   The problem of understanding the origin of the stellar IMF and its seeming universality remains unsolved, but recent work suggests that the mass scale for cloud fragmentation may depend importantly on the thermal physics of collapsing and fragmenting clouds.  Nearly all numerical work has used a simple isothermal equation of state, but if collapsing clouds develop filamentary structures and if most of the star formation occurs in filaments, then the collapse and fragmentation of such configurations can be quite sensitive to variations in temperature.  If, as expected, there is a low-density regime of decreasing temperature and a high-density regime of slowly rising temperature, fragmentation seems likely to be favored at the transition density where the temperature reaches a minimum.  Existing simple treatments of the thermal physics predict that a minimum temperature of about 5~K is attained at a density of about $2 \times 10^{-18}$\,g\,cm$^{-3}$, at which point the Jeans mass is about 0.3 solar masses, close to the mass at which the IMF peaks.  Establishing more clearly the nature of any connection between these quantities will be an important goal of continuing research.

\begin{chapthebibliography}{1}

Abel, T., Bryan, G., \& Norman, M. L. 2000, ApJ, 540, 39

Abel, T., Bryan, G., \& Norman, M. L. 2002, Science, 295, 93

Bate, M. R., Bonnell, I. A., \& Bromm, V. 2003, MNRAS, 339, 577

Bonnell, I. A., \& Bate, M. R., 2002, MNRAS, 336, 659

Bromm, V. 2005, in The Initial Mass Function, ed.\ E. Corbelli, F. Palla,
  \& H. Zinnecker (Dordrecht: Kluwer Acad.\ Publ.), in press (this
  volume)

Bromm, V., \& Larson, R. B. 2004, ARA\&A, 42, 79

Bromm, V., \& Loeb, A. 2004, New Astron, 9, 353

Bromm, V., Coppi, P. S., \& Larson, R. B. 1999, ApJ, 527, L5

Bromm, V., Coppi, P. S., \& Larson, R. B. 2002, ApJ, 564, 23

Curry, C. L. 2002, ApJ, 576, 849

Evans, N. J. 1999, ARA\&A, 37, 311

Evans, N. J., Rawlings, J. M. C., Shirley, Y. L., \& Mundy, L. G. 2001,
  ApJ, 557, 193

Hartmann, L. 2002, ApJ, 578, 914

Jappsen, A.-K., Li, Y., Mac Low, M.-M., \& Klessen, R. S. 2003, BAAS,
  35, 1214

Jappsen, A-K., Klessen, R. S., Larson, R. B., Li, Y., \& Mac Low, M.-M.
  2005, A\&A, in press.

Jeans, J. H. 1902, Phil Trans Roy Soc, 199, 49

Jeans, J. H. 1929, Astronomy and Cosmogony  (Cambridge: Cambridge Univ.\
  Press; reprinted by Dover, New York, 1961).

Johnstone, D., Wilson, C.\,D., Moriarty-Schieven, G., Joncas, G., Smith,
  G., Gregersen, E., \& Fich, M. 2000, ApJ, 545, 327

Johnstone, D., Fich, M., Mitchell, G. F., \& Moriarty-Schieven, G. 2001,
  ApJ, 559, 307

Kawachi, T., \& Hanawa, T. 1998, PASJ, 50, 577

Klessen, R. S. 2001, ApJ, 556 837

Klessen, R. S. 2005, in The Initial Mass Function, ed.\ E. Corbelli, F.
  Palla, \& H. Zinnecker (Dordrecht: Kluwer Acad.\ Publ.), in press (this
  volume)

Klessen, R. S., \& Burkert, A. 2001, ApJ, 549, 386

Koyama, H., \& Inutsuka, S.-I. 2000, ApJ, 532, 980

Kroupa, P. 2002, Science, 295, 82

Lada, C. J., \& Lada, E. A. 2003, ARA\&A, 41, 57

Larson, R. B. 1973, Fundam Cosmic Phys, 1, 1

Larson, R. B. 1978, MNRAS, 184, 69

Larson, R. B. 1985, MNRAS, 214, 379

Larson, R. B. 1996, in The Interplay Between Massive Star Formation,
  the ISM, and Galaxy Evolution, ed.\ D. Kunth, B. Guiderdoni, M.
  Heydari-Malayeri, \& T. X. Thuan (Gif sur Yvette: Editions Fronti\`eres),
  p.\ 3

Larson, R. B. 2003a, in Galactic Star Formation Across the Stellar
  Mass Spectrum, ASP Conf.\ Ser.\ Vol. 287, ed.\ J. M. De Buizer \&
  N. S. van der Bliek (San Francisco: Astron.\ Soc.\ Pacific), p.\ 65

Larson, R. B. 2003b, Rep Prog Phys, 66, 1651

Li, Y., Klessen, R. S., \& Mac Low, M.-M. 2003, ApJ, 592, 975

Low, C., \& Lynden-Bell, D. 1976, MNRAS, 176, 367 

Luhman, K. L., \& Rieke, G. H. 1999, ApJ, 525, 440

Mac Low, M.-M., \& Klessen, R. S. 2004, Rev Mod Phys, 76, 125

Masunaga, H., \& Inutsuka, S.-I. 2000, ApJ, 531, 350

Monaghan, J. J., \& Lattanzio, J. C. 1991, ApJ, 375, 177

Motte, F., \& Andr\'e, P. 2001a, in From Darkness to Light: Origin
  and Evolution of Young Stellar Clusters, ASP Conf.\ Ser.\ Vol.\ 243,
  ed.\ T. Montmerle \& P. Andr\'e (San Francisco: Astron.\ Soc.\ Pacific),
  p.\ 310

Motte, F., \& Andr\'e, P. 2001b, A\&A, 365, 440

Motte, F., Andr\'e, P., \& Neri, R. 1998, A\&A, 336, 150

Myers, P. C. 1978, ApJ, 225, 389

Nakamura, F., \& Umemura, M. 2001, ApJ, 548, 19

Nakamura, F., \& Umemura, M. 2002. Prog Theor Phys Suppl, 147, 99

Omukai, K., \& Palla, F. 2003, ApJ, 589 677

Scalo, J. M. 1986, Fundam Cosmic Phys, 11, 1

Scalo, J. M. 1998, in The Stellar Initial Mass Function, ASP Conf.\
  Ser.\ Vol.\ 142, ed.\ G. Gilmore \& D. Howell (San Francisco: 
  Astron.\ Soc.\ Pacific), p.\ 201

Schneider, S., \& Elmegreen, B. G. 1979, ApJ Suppl, 41, 87

Tafalla, M., Myers, P. C., Caselli, P., \& Walmsley, C. M. 2004, A\&A,
  416, 191

Zucconi, A., Walmsley, C. M., \& Galli, D. 2001, A\&A, 376, 650

\end{chapthebibliography}

\end{document}